\documentclass[11pt,a4paper]{article}
\usepackage{jcappub}

\usepackage{bm}
\usepackage{epsfig}
\usepackage{citesort}
\usepackage{graphicx}
\usepackage{color}

\newcommand{\lwig}{\mbox{\;\raisebox{.3ex}
    {$<$}$\!\!\!\!\!$\raisebox{-.9ex}{$\sim$}\;}}

\newcommand{\lambdabar}%
{{\hbox{$\lambda$\kern-1.ex\raise+0.45ex\hbox{--}}}}

\begin{document}


\subheader{\hfill MPP-2011-100\\ \hbox{\  } \hfill TTK-11-29}

\title{Sterile neutrinos with eV masses in cosmology -- how disfavoured
exactly?}

\author[a]{Jan Hamann,}
\author[a]{Steen Hannestad,}
\author[b]{Georg G.~Raffelt}
\author[c]{and Yvonne~Y.~Y.~Wong}

\affiliation[a]{Department of Physics and Astronomy\\
 University of Aarhus, DK-8000 Aarhus C, Denmark}

\affiliation[b]{Max-Planck-Institut f\"Ru Physik (Werner-Heisenberg-Institut)\\
 F\"ohringer Ring 6, D-80805 M\"unchen, Germany}

\affiliation[c]{Institut f\"ur Theoretische Teilchenphysik und Kosmologie\\
 RWTH Aachen, D-52056 Aachen, Germany}

\emailAdd{hamann@phys.au.dk}
\emailAdd{sth@phys.au.dk}
\emailAdd{raffelt@mppmu.mpg.de}
\emailAdd{yvonne.wong@physik.rwth-aachen.de}

\abstract{We study cosmological models that contain sterile
neutrinos with eV-range masses as suggested by reactor and
short-baseline oscillation data. We confront these models with 
 both precision cosmological
data (probing the CMB decoupling epoch) and light-element abundances
(probing the BBN epoch).  
In the minimal $\Lambda$CDM model, such sterile
neutrinos are strongly disfavoured by current data because they
contribute too much hot dark matter.
However, if the cosmological framework is extended to include also
additional relativistic degrees of freedom beyond the three standard neutrinos 
and the putative sterile neutrinos,
then the hot dark matter constraint on the sterile states is considerably relaxed.
 A further improvement is achieved by
allowing a dark energy equation of state parameter $w<-1$. While BBN
strongly disfavours extra radiation beyond the assumed eV-mass
sterile neutrino, this constraint can be circumvented by  a small
$\nu_e$ degeneracy. Any model containing eV-mass sterile neutrinos
implies also strong modifications of other cosmological parameters.
Notably,  the inferred cold dark matter density can shift up by 20--75\% relative
to the standard $\Lambda$CDM value.}

\maketitle

\section{Introduction}                        \label{sec:introduction}

There is mounting evidence from reactor and short-baseline neutrino
oscillation experiments suggesting the existence of one or two
sterile neutrinos with mass splittings relative to the active flavours in
the neighbourhood of $\Delta m^2 \sim 1~{\rm eV}^2$ and fairly large
mixing  parameters (see, e.g.,~\cite{Kopp:2011qd}
for a review and global interpretation, and~\cite{Akhmedov:2010vy,Agarwalla:2010zu,Giunti:2011gz} for
other recent analyses).  In the early universe, flavour oscillations
would bring these sterile states into thermal equilibrium prior to
neutrino decoupling at $T\sim 1$~MeV, thereby increasing the
relativistic energy density (see, e.g.,~\cite{Dolgov:2002wy} for a review). This increase, commonly parameterised
in terms of an additional contribution to the relativistic neutrino
degrees of freedom $N_{\rm eff}$, significantly modifies big-bang
nucleosynthesis (BBN) of light elements, the cosmic microwave
background (CMB) anisotropies, and the formation of large-scale
structures (LSS). By the same token, these eV-mass sterile neutrinos
would later play the role of a non-negligible hot-dark matter
component.

The cosmological verdict on eV-mass sterile neutrinos is somewhat
mixed. Current CMB and LSS observations show a consistent
preference for additional relativistic degrees of freedom beyond the
standard model expectation of $N_{\rm eff}=3.046$, with 
low to moderate statistical
significance~\cite{Izotov:2010ca,Hamann:2010bk,Giusarma:2011ex,Hou:2011ec,Keisler:2011aw,GonzalezMorales:2011ty}.
On the other hand, if this putative radiation excess is interpreted
in terms of sterile neutrinos, the usual hot dark matter limits
constrain the sterile masses to the sub-eV regime. For example, in a
3+1 scenario consisting of three essentially massless active
flavours and one fully thermalised sterile flavour, the sterile mass
is constrained to $m_s \lwig 0.5$~eV \cite{Hamann:2010bk}.  Thus,
cosmology appears to disfavour the existence of one or two sterile
states that have mass and mixing parameters  favoured by a global analysis
of the laboratory data.

However, both the tentative evidence for extra radiation and these
hot-dark matter bounds are based on the simplest $\Lambda$CDM
framework; extended models may provide different answers.
For example, if the dark energy equation of state parameter $w$ and
curvature $\Omega_k$ are used as fit parameters, the presence of
eV-mass sterile neutrinos favours a $w$ value much smaller than $-1$
\cite{Kristiansen:2011mp}. Here, we take the view that it may be more
natural to look for extensions to $\Lambda$CDM  within the neutrino
sector alone, for example in terms of additional radiation.
(However, we note in passing that in certain dynamical dark energy
models such as the mass varying neutrino scenario, a strongly
modified neutrino sector is responsible for driving the dark energy
evolution. In this sense, changing $w$ may not be an entirely
unnatural solution to the massive sterile neutrino problem.)
Multiple sterile right-handed states may exist and some of them may
even be lighter that the eV range. Moreover, the existence of
sterile neutrinos also provides a means for the generation of large
neutrino chemical potentials~\cite{Foot:1995qk}, which may be exploited to circumvent
standard BBN bounds on the number of relativistic degrees of
freedom.

The premise of our work is that sterile neutrino states with
eV-range masses are real and that they are fully thermalised prior
to neutrino decoupling. In addition, we allow for radiation beyond
that due to the three standard flavours (assumed to be
effectively massless) and beyond the eV-mass sterile neutrinos. We
consider both data from precision cosmology and the constraints on
$N_{\rm eff}$ imposed by BBN.

More specifically, in section~\ref{sec:cmblss} we investigate the
impact of eV-mass sterile neutrinos on the inference of cosmological
parameters from CMB and LSS observations. In section \ref{sec:bbn}
we evaluate the constraints on additional relativistic species
during the BBN epoch, using measurements of the primordial
abundances of several light elements.
In section \ref{sec:discussion} we summarise and discuss our results.

\section{CMB and LSS\label{sec:cmblss}}

\subsection{Cosmological data\label{sec:data}}

The CMB anisotropies and the LSS distribution are sensitive to both
additional relativistic degrees of freedom and the mass scale of
light, free-streaming particles. To explore the impact of eV-mass
sterile neutrinos on precision cosmology, we use CMB anisotropy data
and their accompanying likelihood routines from the WMAP 7-year data
release~\cite{Komatsu:2010fb}, as well as the
ACBAR~\cite{Reichardt:2008ay}, BICEP~\cite{Chiang:2009xsa}, and
QuAD~\cite{Brown:2009uy} experiments.  In addition, we use the halo
power spectrum extracted from the SDSS-DR7 luminous red galaxy
sample~\cite{Reid:2009xm}, and type Ia supernova (SN) data from the
Union-2 compilation~\cite{Amanullah:2010vv}. Finally, we impose a
constraint on the Hubble parameter based on the Hubble Space
Telescope observations~\cite{Riess:2009pu}. We refrain from using
observational data pertaining to very small length scales, such as
estimates of small-scale density fluctuation amplitudes from the
Lyman-$\alpha$ forest and cluster abundance.  While these
measurements are in principle an extremely powerful tool for
constraining neutrino masses, they are also currently dominated by
systematic uncertainties.  We have tested explicitly that adding
data from the Atacama Cosmology Telescope (ACT)~\cite{Hlozek:2011pc}
does not alter our results;   The main role of ACT is to break 
parameter degeneracies in CMB-only analyses.  These degeneracies are however 
already broken by the addition of LSS data.

We construct allowed regions in cosmological parameter space using
standard Bayesian inference techniques.  All posterior probability
density functions are sampled using  the Markov Chain Monte Carlo
(MCMC) code \texttt{CosmoMC}~\cite{Lewis:2002ah}.

\begin{table}[th]
\caption{Definitions of the model parameters and the prior ranges
imposed on them in our CMB+LSS likelihood analysis.  Uniform priors
are used for all listed parameters. \label{tab:prior}}
\begin{center}
{\footnotesize
\begin{tabular}{|ll|c|}
\hline
Parameter   &  Symbol & Prior  \\
\hline
Baryon density &$\omega_{\rm b}$ & $0.005 \to 0.1$\\
Cold dark matter density & $\omega_{\rm cdm}$ & $0.01 \to 0.99$\\
Hubble parameter & $h$ & $0.4 \to 1.0$ \\
Amplitude of scalar spectrum @ $k=0.05~{\rm Mpc}^{-1}$ &  $\log[10^{10}A_s]$ & $2.7 \to 4$\\
Scalar spectral index & $n_{\rm s}$ & $0.5 \to 1.5$\\
Optical depth to reionisation & $\tau$ & $0.01 \to 0.8$ \\
\hline
Number of extra massless neutrino degrees of freedom & $\Delta N_{\rm ml}$ & $0 \to 5$ \\
Dark energy equation of state parameter & $w$ & $-2.5 \to -0.5$ \\
\hline
\end{tabular}
}
\end{center}
\end{table}

\subsection{Cosmological models\label{sec:models}}

We consider three basic cosmological frameworks in which we embed
light sterile neutrinos and analyse their consequences.

\begin{enumerate}
\item  The $\Lambda$CDM class of models is defined by a flat
    spatial geometry and  six free parameters, $\{\omega_{\rm
    b},\omega_{\rm cdm},h,A_s,n_s,\tau\}$. See
    table~\ref{tab:prior} for their definitions and prior
    ranges. Within this framework, we consider four possibilities in
    the neutrino sector: (a) 3 massless neutrinos (standard
    $\Lambda$CDM), (b) 3 massless+1 sterile (0~eV), (c)
    3~massless+1~sterile (1~eV), and (d) 3~massless+1 sterile
    (2~eV).

\item  In the second class of models, which we dub
    $\Lambda$CDM+$\Delta N$, we include additional relativistic degrees
    of freedom beyond the 3+1 standard and sterile neutrinos. We
    consider three scenarios: (a) $3+\Delta N_{\rm
    ml}$~massless+1~sterile (0~eV), (b) $3+\Delta N_{\rm
    ml}$~massless+1~sterile (1~eV), and (c) $3+ \Delta N_{\rm
    ml}$~massless+1~sterile (2~eV). Such models are especially
    sensitive to BBN constraints to be discussed in
    section~\ref{sec:bbn}.

\item The $w$CDM+$\Delta N$ framework is a variant of $\Lambda$CDM+$\Delta N$,
    in which we extend the dark energy sector to include the
    possibility that $w \neq -1$.

\end{enumerate}

\subsection{Goodness-of-fit\label{sec:gof}}

We consider first the goodness-of-fit, as quantified by the best-fit
effective $\chi^2$, of the cosmological models described in the
previous section.  Here the effective $\chi^2$ is defined as
$\chi^2_{\rm eff} = - 2 \ln \mathcal{L}_{\rm max}$, where $
\mathcal{L}_{\rm max}$ is the maximum likelihood of the data given the model.
Table~\ref{tab:chi2} summarises the best-fit $\chi^2_{\rm eff}$ for
these models and, where appropriate, the preferred values of $\Delta N_{\rm
  ml}$ and $w$,  using the data sets described in
section~\ref{sec:data}. Some comments are in order.

\begin{enumerate}

\item Within the $\Lambda$CDM framework, a scenario with 3 massless
  neutrinos plus one fully thermalised massless sterile species offers
  a slightly better fit than standard  $\Lambda$CDM ($\Delta \chi^2_{\rm
    eff} = - 3.16$). However, once the sterile states have mass, the
  quality of the fit deteriorates.  The 1~eV scenario is already
  marginally worse than standard $\Lambda$CDM by $\Delta \chi^2_{\rm
    eff} =4.20$. The 2~eV model may be deemed unacceptable
  ($\Delta\chi^2_{\rm eff} = 21.41$).

\item The situation improves when we allow for additional
    radiation ($\Lambda$CDM+$\Delta N$). For example, if the
    sterile neutrino has a mass of 1~eV, the best-fit 
    $\chi^2_{\rm eff}$ is comparable to that of standard
    $\Lambda$CDM, albeit at the cost of admitting
    $\Delta N_{\rm ml} \sim 1.5$ additional massless degrees of
    freedom. For a 2~eV sterile
     mass, we find $\Delta N_{\rm ml} \sim 2.6$ and $\chi^2_{\rm
    eff}=12.8$.   This last result can be compared with the 
    $w$CDM+$\Omega_k$ model of reference~\cite{Kristiansen:2011mp}, for which 
     $\Delta \chi^2_{\rm eff} = 12$ assuming a {\em lighter}  1.33~eV sterile neutrino.
    Therefore, introducing extra radiation appears to be somewhat superior to modifying the 
    dark energy sector at resolving the sterile mass
    conundrum.

\item Even more improvement is available if, in addition, we allow the
  dark energy equation of state parameter $w$ to differ from $-1$
  ($w$CDM+$\Delta N$).  In this class of models, we see that a
  scenario with one species of 1~eV sterile neutrinos in fact provides
  a better fit to the data than does standard $\Lambda$CDM,
  with $\Delta \chi^2_{\rm eff} =-0.78$, at the expense of two
  additional free parameters.

\end{enumerate}

\begin{table}[th]
  \caption{Best-fit $\Delta \chi^2_{\rm eff}$ relative to the standard $\Lambda$CDM framework for the models described in section~\ref{sec:models}, using the data sets of section~\ref{sec:data}.  We also show the best-fit values and 95\%-credible upper and lower limits on $\omega_{\rm cdm}$, and, where appropriate, on $\Delta N_{\rm ml}$ and $w$.
  \label{tab:chi2}}
\begin{center}
{\footnotesize
\begin{tabular}{|ll|c|ccc|}
\hline
Framework                  & Neutrino sector & $\Delta \chi^2_{\rm eff}$ & $\Delta N_{\rm ml}$ & $w$ & $\omega_{\rm cdm}$\\ \hline
$\Lambda$CDM         & 3 massless   & $0$ & -- & -- & $0.1132^{+0.0036}_{-0.0082}$ \\
 	&3 massless + 1 sterile (0 eV)     & $-3.16$ & -- & -- & $0.1299^{+0.0069}_{-0.0066}$\\
& 3 massless + 1 sterile (1 eV)      & $4.20$  & -- & -- & $0.1398^{+0.0061}_{-0.0074}$ \\
&  3 massless + 1 sterile (2 eV)       & $21.41$ & -- & -- & $0.1473^{+0.0075}_{-0.0064}$\\
\hline
$\Lambda$CDM+$\Delta N$ & 3+$\Delta N_{\rm ml}$ massless + 1 sterile (0 eV)    & $ -3.54$ & $0.01^{+1.12}_{-0.01}$ & -- & $0.133^{+0.023}_{-0.005}$ \\
& 3+$\Delta N_{\rm ml}$ massless + 1 sterile (1 eV)      & $2.26$ & $1.49^{+1.11}_{-0.73}$ & -- & $0.166^{+0.026}_{-0.017}$ \\
& 3+$\Delta N_{\rm ml}$ massless + 1 sterile (2 eV)      &  $12.82$ & $2.57^{+1.24}_{-0.59}$ & -- & $0.192^{+0.031}_{-0.015}$ \\
\hline
$w$CDM+$\Delta N$  & 3+$\Delta N_{\rm ml}$ massless + 1 sterile (0 eV)      & $-5.38$ & $0.09^{+1.61}_{-0.09}$ & $-1.00^{+0.18}_{-0.12}$ & $0.132^{+0.032}_{-0.006}$\\
& 3+$\Delta N_{\rm ml}$ massless + 1 sterile (1 eV)     & $-0.78$ & $1.23^{+1.61}_{-0.75}$ & $-1.11^{+0.18}_{-0.21}$ & $0.164^{+0.035}_{-0.015}$\\
& 3+$\Delta N_{\rm ml}$ massless + 1 sterile (2 eV)    & $7.80$ & $2.48^{+1.71}_{-0.79}$ & $-1.17^{+0.23}_{-0.22}$ & $0.198^{+0.032}_{-0.019}$ \\
\hline
\end{tabular}
}
\end{center}
\end{table}

\subsection{Effects on other cosmological parameters}

We have seen that precision cosmological  observations can
reasonably accommodate one fully thermalised species of massive
sterile neutrinos if we allow also for additional massless degrees
of freedom and/or a non-standard dark energy equation of state.
An interesting consequence is that the preferred values of other
free parameters of the model also shift accordingly.

\begin{figure}[th]
\center
\includegraphics[height=0.48\textwidth,angle=270]{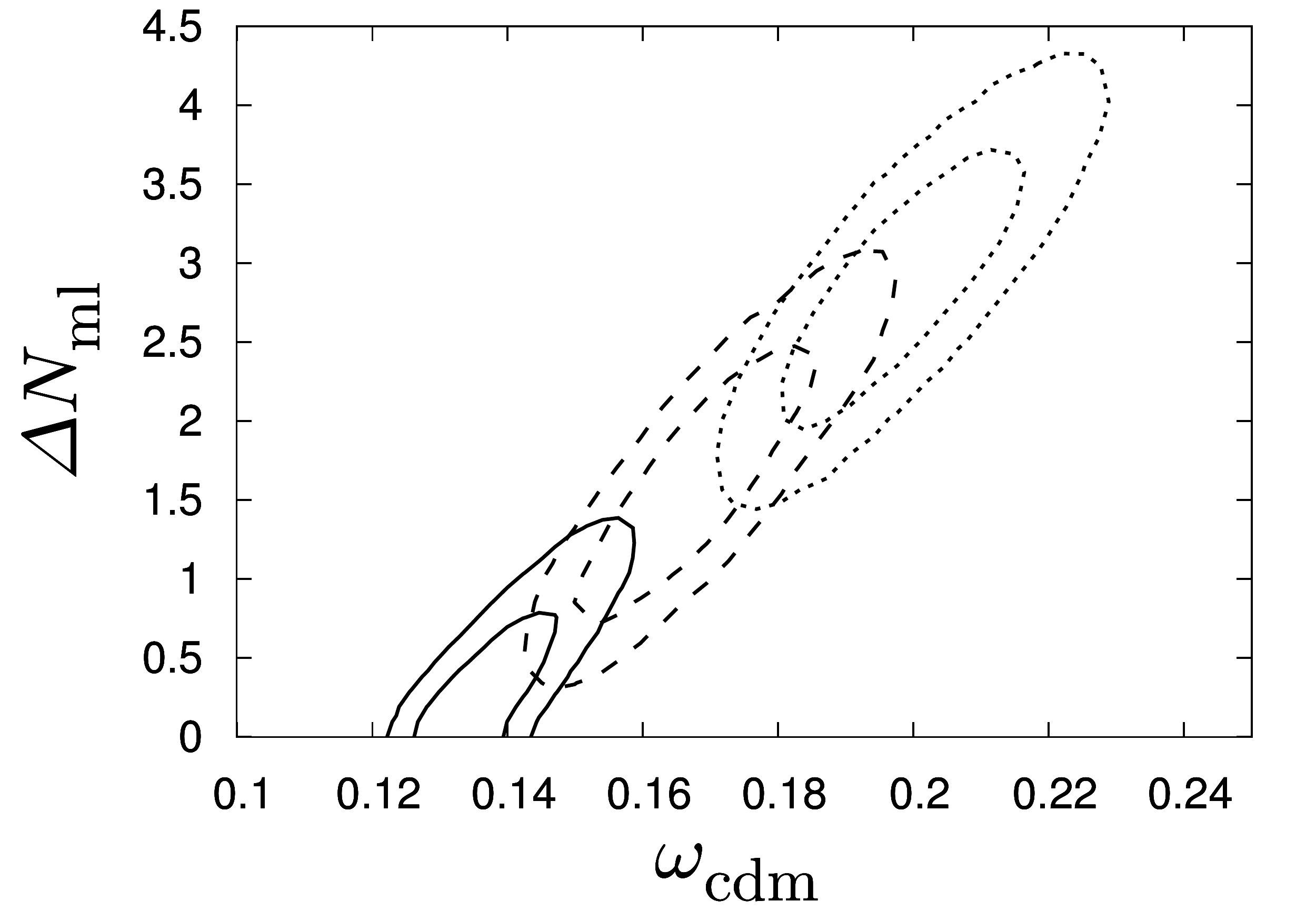}
\caption{2D marginal 68\%- and 95\%-credible regions in the $(\Delta
  N_{\rm ml},\omega_{\rm cdm})$-plane for three $\Lambda$CDM+$\Delta N$ class
  models containing one thermalised sterile species of mass $m_{\rm s} =
  0$~eV (solid), 1~eV (dashed), and 2~eV (dotted).\label{fig:n-dm}}
\end{figure}

\begin{figure}[th]
\center
\includegraphics[height=0.48\textwidth,angle=270]{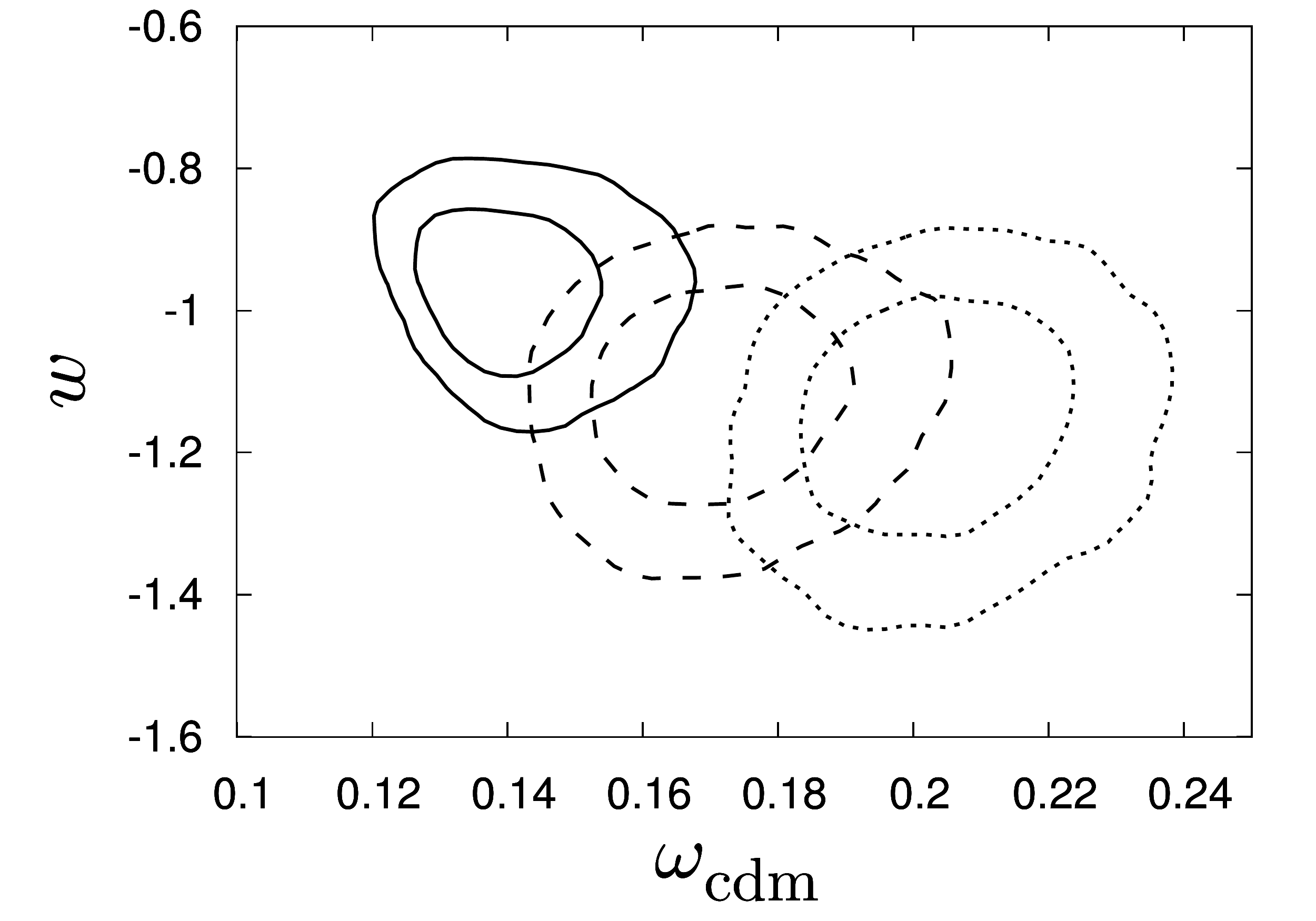}
\caption{2D marginal 68\%- and 95\%-credible regions in the
  $(w,\omega_{\rm cdm})$-plane for three $w$CDM+$\Delta N$ class models containing
  one thermalised sterile species of mass $m_{\rm s} = 0$~eV (solid),
  1~eV (dashed), and 2~eV (dotted). \label{fig:w-dm}}
\end{figure}

The most notable example in this regard is the cold dark matter density
$\omega_{\rm cdm}$.  Figure~\ref{fig:n-dm} illustrates the shift in
$\omega_{\rm cdm}$ as a function of the sterile neutrino mass within
the $\Lambda$CDM+$\Delta N$ framework.  Figure~\ref{fig:w-dm} is similar,
but for the $w$CDM+$\Delta N$ models.  See also table~\ref{tab:chi2} for the best-fit values
and credible regions.  Clearly, the larger the sterile
neutrino mass,  the larger the preferred value of $\omega_{\rm cdm}$.
In the case of a 2~eV sterile neutrino, the upward shift in $\omega_{\rm cdm}$ can be as large as
75\% in the $w$CDM+$\Delta N$ model, relative to the standard $\Lambda$CDM inferred value.
This shift in the cold dark matter density  can have importance
consequences for, e.g., the SUSY dark matter parameter space.

Another affected parameter is the scalar spectral index $n_s$, whose
preferred region widens in the presence of additional light species,
as was also seen in our previous analysis \cite{Hamann:2010bk}.

\section{Big bang nucleosynthesis \label{sec:bbn}}

While not sensitive to the masses of neutrinos, BBN has long been
used to probe the radiation content of the Universe at temperatures
of order 1~MeV~\cite{Shvartsman:1969mm,Steigman:1977kc,Simha:2008zj}. In this
section, we explore the implications of the latest primordial
element abundance measurements on the sterile neutrino scenario.  We
consider a general BBN model with three free parameters: the baryon
density $\omega_{\rm b}$, an extra $N_{\rm s}$ effective sterile
neutrino species on top of the usual three fully thermalised standard
neutrinos, and, eventually we also allow for the presence of a
neutrino chemical potential $\xi$.  Table~\ref{tab:bbnprior}
summarises the prior ranges for these parameters.

\begin{table}[ht]
\caption{Definitions of the model parameters and the prior ranges
imposed on them in our BBN likelihood analysis.  Uniform priors are
used for all listed parameters. \label{tab:bbnprior}}
\begin{center}
{\footnotesize
\begin{tabular}{|ll|c|}
\hline
Parameter   &  Symbol & Prior  \\
\hline
Baryon density &$\omega_{\rm b}$ & $0.01 \to 0.035$\\
Number of sterile neutrinos & $N_{\rm s}$ & $0 \to 5$ \\
Neutrino chemical potential & $\xi$ & $-0.2 \to 0.3$ \\
\hline
\end{tabular}
}
\end{center}
\end{table}

\subsection{Analysis}

\begin{figure}[t]
\center
\includegraphics[height=.48\textwidth,angle=270]{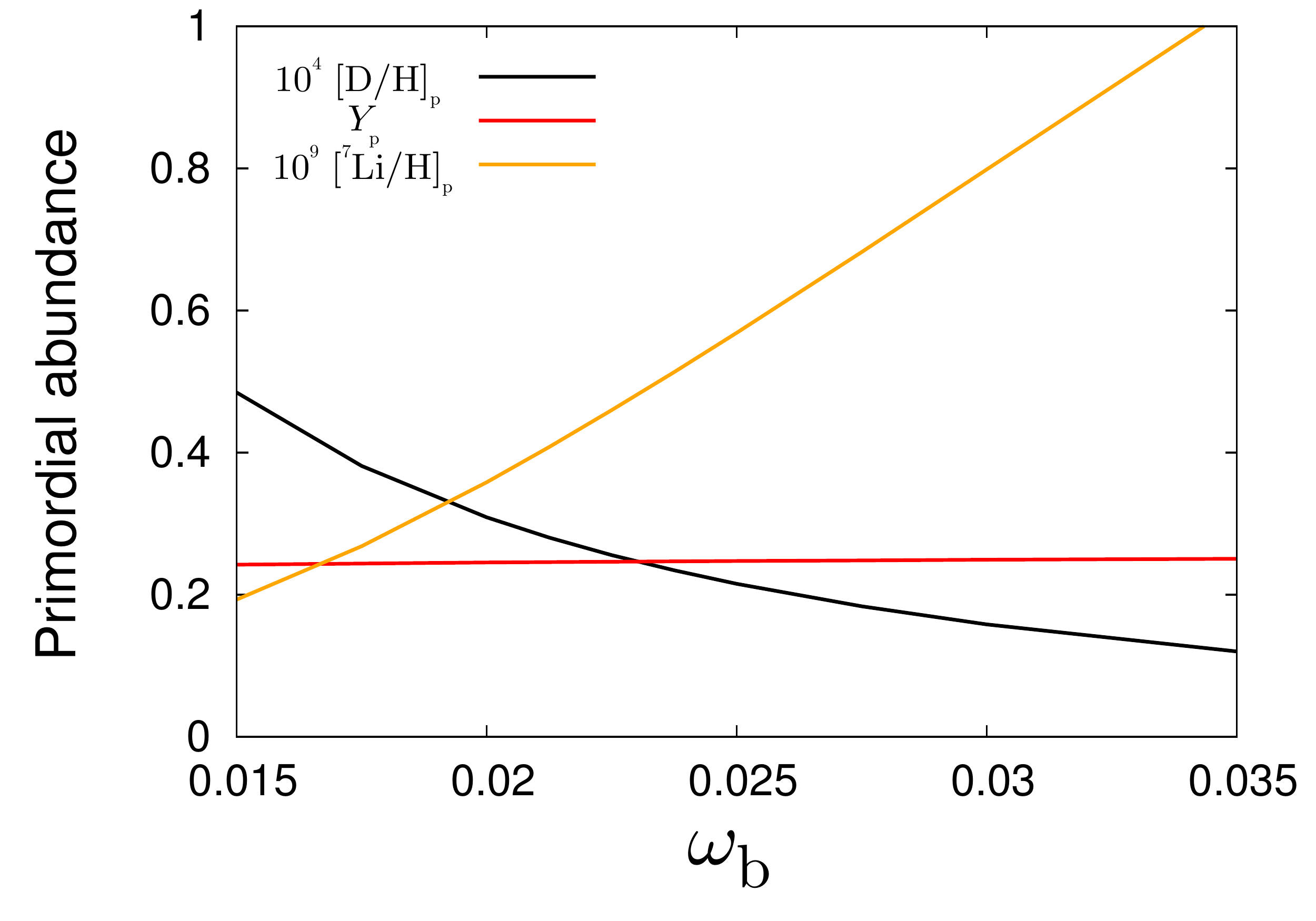}
\includegraphics[height=.48\textwidth,angle=270]{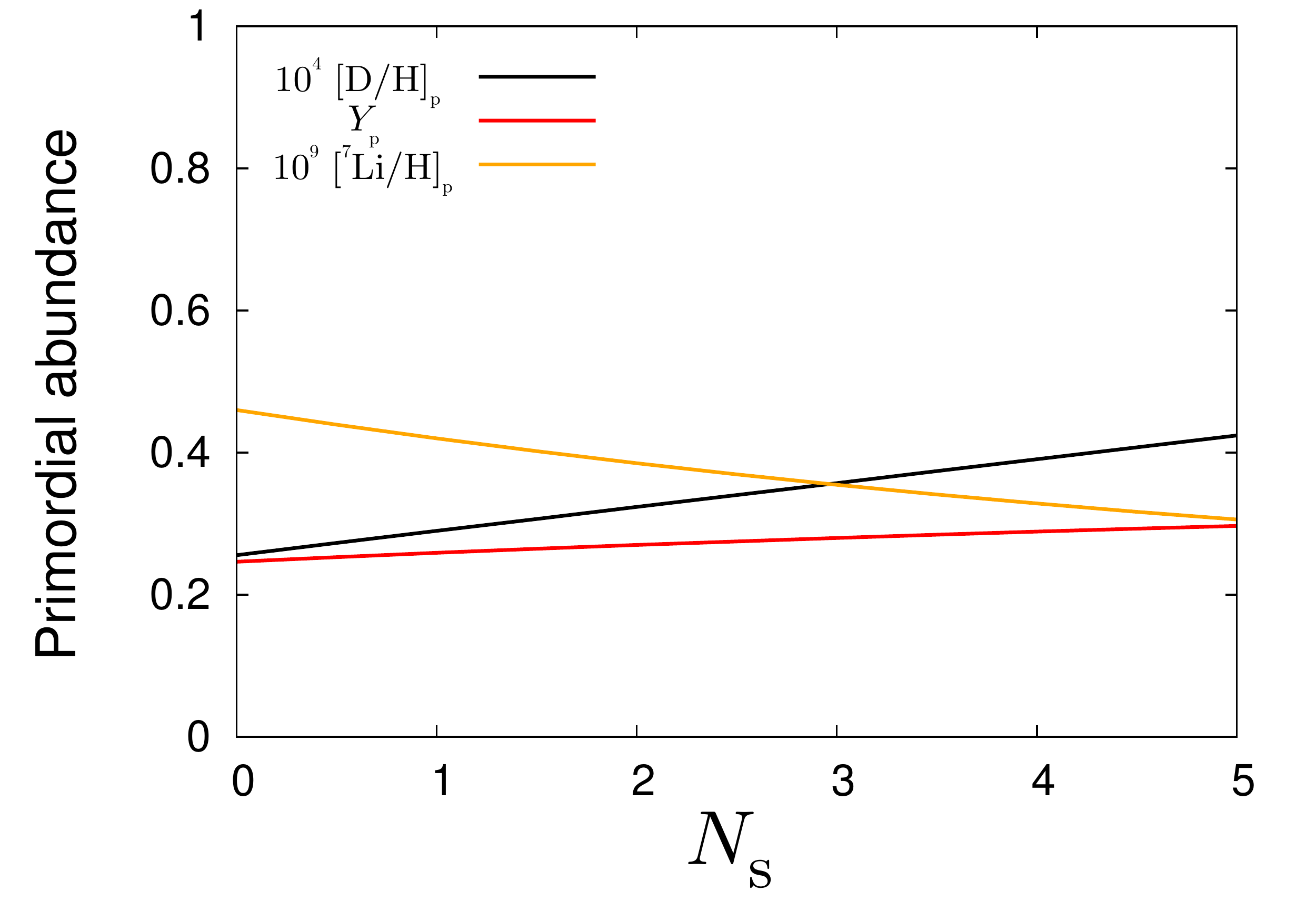}
\begin{center}
\includegraphics[height=.48\textwidth,angle=270]{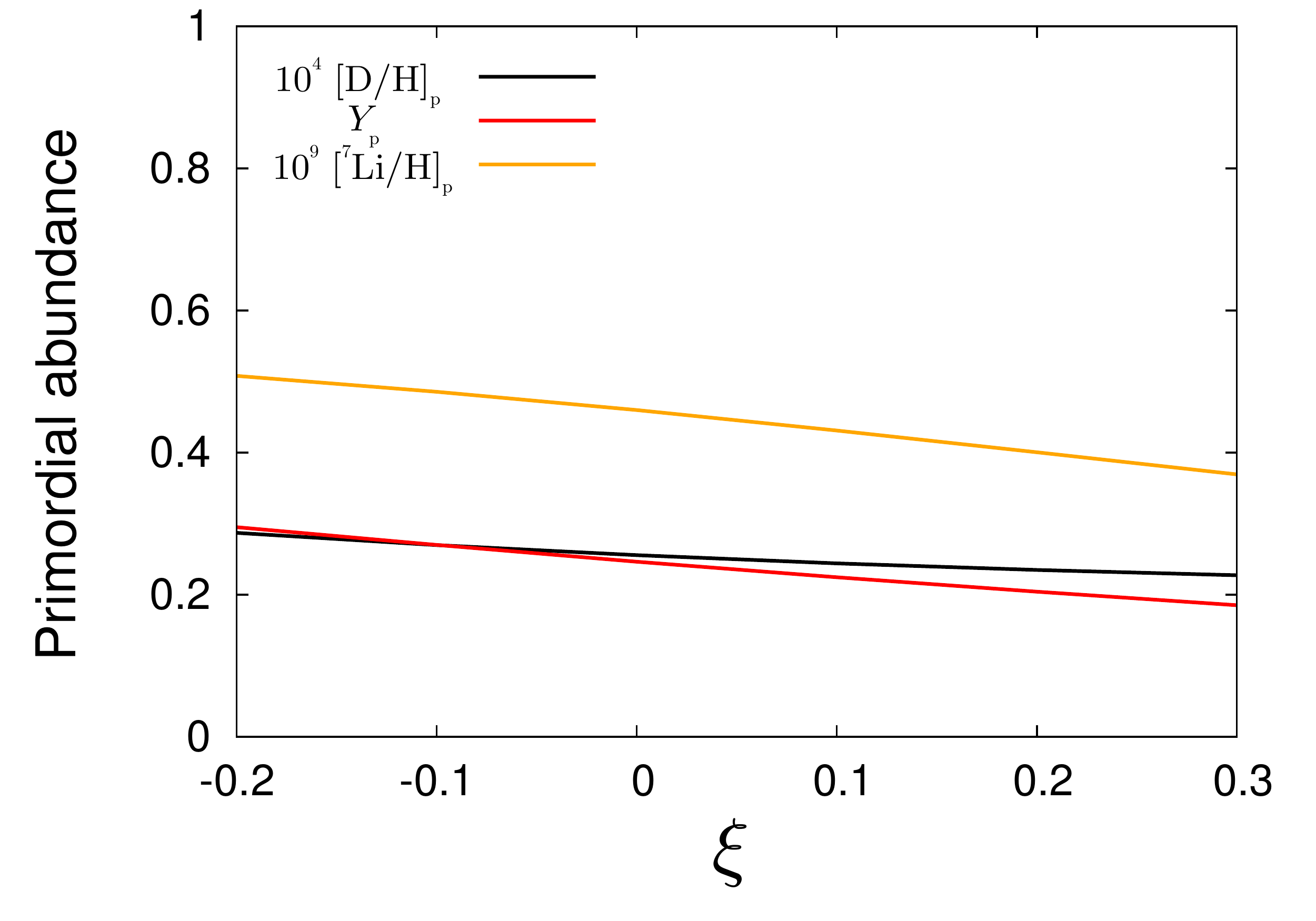}
\end{center}
\caption{Primordial abundances of D, $^4$He and $^7$Li as functions of the cosmological parameters.  {\it Top left:} dependence on $\omega_{\rm b}$ for $N_{\rm s} = \xi = 0$. {\it Top right:} dependence on $N_{\rm s}$ for $\omega_{\rm b} = 0.0225$ and $\xi = 0$.  {\it Bottom:} dependence on $\xi$ for $\omega_{\rm b} = 0.0225$ and $N_{\rm s} = 0$.
  \label{fig:bbn0}}
\end{figure}

We infer constraints on the BBN parameters with a modified version
of the Markov Chain Monte Carlo (MCMC)
sampler~\texttt{CosmoMC}~\cite{Lewis:2002ah}.%
\footnote{The numerical
  module for implementing the BBN likelihood in \texttt{CosmoMC} is
  available from the corresponding author (Jan Hamann) upon request.}
The \texttt{PArthENoPE}~\cite{Pisanti:2007hk} code is employed to
pre-calculate the primordial element abundances on a grid of points
in $(\omega_{\rm b},N_{\rm s},\xi)$-space.  For each set of
parameter values the abundances are then obtained from the grid by
3-dimensional cubic spline interpolation.  We illustrate the
dependence of the light element abundances on the cosmological
parameters on selected slices through parameter space in
figure~\ref{fig:bbn0}.

\subsection{Uncertainties of input data}

The output of \texttt{PArthENoPE} is subject to theoretical
uncertainties from two principal sources: the nuclear reaction rates
and the free neutron lifetime $\tau_{\rm n}$. The former induce a
$\pm1.6$\% error for the D prediction, negligible error for $^4$He,
and $\pm8$\% for  $^7$Li~\cite{Serpico:2004gx}. We fold these uncertainties into our
definitions of the corresponding likelihood functions in
section~\ref{sec:like}.

The main source of uncertainty for the predicted $^4$He abundance is the free
neutron lifetime. The Particle Data Group recommends a value of
$\tau_{\rm n}^{\rm PDG} = 885.7 \pm 0.8$~s~\cite{Nakamura:2010zzi}.
This, however, appears to be in strong tension with recent
measurements performed by Serebrov~{\it et
  al.}~\cite{Serebrov:2004zf}, who found $\tau_{\rm n}^{\rm S} = 878.5
\pm 0.8$~s, and Pichlmair~{\it et al.}~\cite{Pichlmaier:2010zz}, who
measured \mbox{$\tau_{\rm n}^{\rm P} = 880.7 \pm 1.8$~s}.  Although a
re-analysis of neutron lifetime data shows that previous estimates may
have been biased by $\sim +6$~s~\cite{Serebrov:2010sg}, we will
nonetheless consider the two extreme values $\tau_{\rm n}^{\rm S}$
and $\tau_{\rm n}^{\rm PDG}$ in the following, in order to illustrate
the possible impact of this uncertainty on our inference. Going from
$\tau_{\rm n}^{\rm S}$ to $\tau_{\rm n}^{\rm PDG}$ shifts the $Y_{\rm
  p}$ prediction by $\sim 0.6\%$, whereas $\left[^2{\rm D}/{\rm
    H}\right]_{\rm p}$ and $\left[^7{\rm Li}/{\rm H}\right]_{\rm p}$
change by $\sim 0.3\%$ which is negligible compared with the effects of other
uncertainties (nuclear rates and astrophysical measurements).

\subsection{Primordial element abundance data and likelihood functions{\label{sec:like}}}
\subsubsection*{Deuterium}
The primordial deuterium abundance $\left[{\rm D}/{\rm H}
\right]_{\rm p}$ can be inferred from measurements of the absorption
of quasar light in high-redshift low-metallicity hydrogen clouds.
Pettini {\it et al.}~\cite{Pettini:2008mq} derive a value of $\log
\left[{\rm D}/{\rm H} \right]_{\rm p} = -4.55 \pm 0.03$ from the
observation of the spectra of seven quasars.  Taking into account
the theoretical uncertainty of 1.6\%, the corresponding likelihood
function is
\begin{equation}
  -2 \ln \mathcal{L}_{\rm D} = \frac{\left(\log \left[ {\rm D}/{\rm H} \right]_{\rm p} + 4.55 \right)^2}{0.034^2}.
\end{equation}
\subsubsection*{Helium-4}
The mass fraction $Y$ of $^4$He is determined by measuring helium
and hydrogen emission lines in low-metallicity HII regions of dwarf
galaxies.  Its primordial value $Y_{\rm p}$ can then be estimated by
treating $Y$ as a function of metallicity $Z$ and extrapolating to
$Z=0$.  A linear regression of data from seven high-quality objects
with a careful treatment of systematic uncertainties yields $Y_{\rm
p} = 0.2573^{+0.0033}_{-0.0088}$, with the additional restriction to
theoretically meaningful positive
slopes ${\rm d}Y/{\rm d}Z > 0$ \cite{Aver:2010wd}.%
\footnote{It has been argued that a linear regression may not
  necessarily be realistic in view of certain models of early $^4$He
  production by Pop III stars~\cite{Mangano:2011ar}, as for instance suggested in
  references~\cite{Salvaterra:2003nu,Vangioni:2010wf}.  However, even in
  these cases, the difference between the primordial and the observed values
  of $Y$ will not exceed $\sim 0.01$, a possibility that is well
  covered by the lower limit on $Y_{\rm p}$ in equation~(\ref{eq:helike}).}
We therefore model the likelihood function as
\begin{equation}
\label{eq:helike}
  -2 \ln \mathcal{L}_{\rm ^4He} =\left\{
    \begin{array}{cl}
      \frac{\left( Y_{\rm p} - 0.2573 \right )^2}{0.0033^2},&
     {\rm if}\ Y_{\rm p} \geq 0.2573 ,
     \\  \frac{\left( Y_{\rm p} - 0.2573 \right)^2}{0.0088^2}, &{\rm if}\ Y_{\rm p} < 0.2573.
    \end{array}
\right.
\end{equation}
\subsubsection*{Lithium-7}
Lithium abundances have been determined from the spectra of
metal-poor dwarf stars.  At low metallicities, the lithium abundance
appears to be independent of metallicity~\cite{Spite:1982dd}.
However, the measurements are subject to significant systematic
uncertainties about the stars' temperatures.  Additionally, the
measured plateau value may not be representative of the primordial
abundance: $^7$Li can be generated by cosmic rays, or depleted
through Population III stars or diffusion processes within the dwarf
stars~\cite{Korn:2006tv}.

The Particle Data Group estimates an average of $[^7{\rm Li}/{\rm
  H}]_{\rm p} = (1.7 \pm 0.06 \pm 0.44) \times 10^{-10}$, which is
significantly lower than the standard BBN expectation for realistic
baryon densities -- the well-known lithium
problem~\cite{Cyburt:2008kw}.  Approximating the 8\% theoretical
uncertainty with an absolute error of $\sigma^{\rm nucl}_{^7{\rm
Li}} \simeq 0.4 \times 10^{-10}$ and adding the errors in
quadrature, we arrive at the following likelihood function for
lithium:
\begin{equation}
  -2 \ln \mathcal{L}_{^7{\rm Li}} = \frac{\left( 10^{10} [^7{\rm Li}/{\rm H}]_{\rm p} - 1.7 \right)^2}{0.598^2},
\end{equation}
which will be considered only where appropriate.

\subsubsection*{CMB+LSS prior on the baryon density}
In addition to primordial abundance measurements, we also consider a
prior on the baryon density $\omega_{\rm b}$ from CMB+LSS data as an
independent constraint.  Since the bounds on $\omega_{\rm b}$ are in
principle model-dependent, using results from a fit to the standard
$\Lambda$CDM model as a prior for extended models would be
technically  incorrect. Instead we use the baryon density inferred
within the $\Lambda$CDM+$N_{\rm s}$ model, $\omega_{\rm b} = 0.02255
\pm 0.00049$, to define our prior
\begin{equation}
  -2 \ln \pi_{\rm CMB+LSS}(\omega_{\rm b}) \propto  \frac{\left( \omega_{\rm b} - 0.02255 \right)^2}{0.00049^2}.
\end{equation}
In practice, however, present CMB+LSS data are sufficiently
sensitive to break any potential degeneracy between $N_{\rm s}$ and
$\omega_{\rm b}$, so that  the constraints on $\omega_{\rm b}$  show
no appreciable variation with respect to the inclusion or otherwise
of  the parameter $N_{\rm s}$.

\subsection{BBN with sterile neutrinos}

We first consider a BBN scenario with $N_{\rm s}$ effective sterile
neutrino species.  The additional radiation energy density
contributed by the sterile neutrinos increases the expansion rate at
BBN, leading to a higher freeze-out temperature $T_{\rm f}$.  As a
consequence, the equilibrium neutron-to-proton ratio $n/p = \exp (-
\Delta m/T)$, where $\Delta m$ is the neutron-proton mass
difference, is larger at neutron freeze-out. The resulting $Y_{\rm
p}$ is larger because almost all neutrons end up in $^4$He.
Similarly, a larger neutron lifetime also increases the expected
$Y_{\rm p}$.

\begin{figure}[th]
\includegraphics[height=.48\textwidth,angle=270]{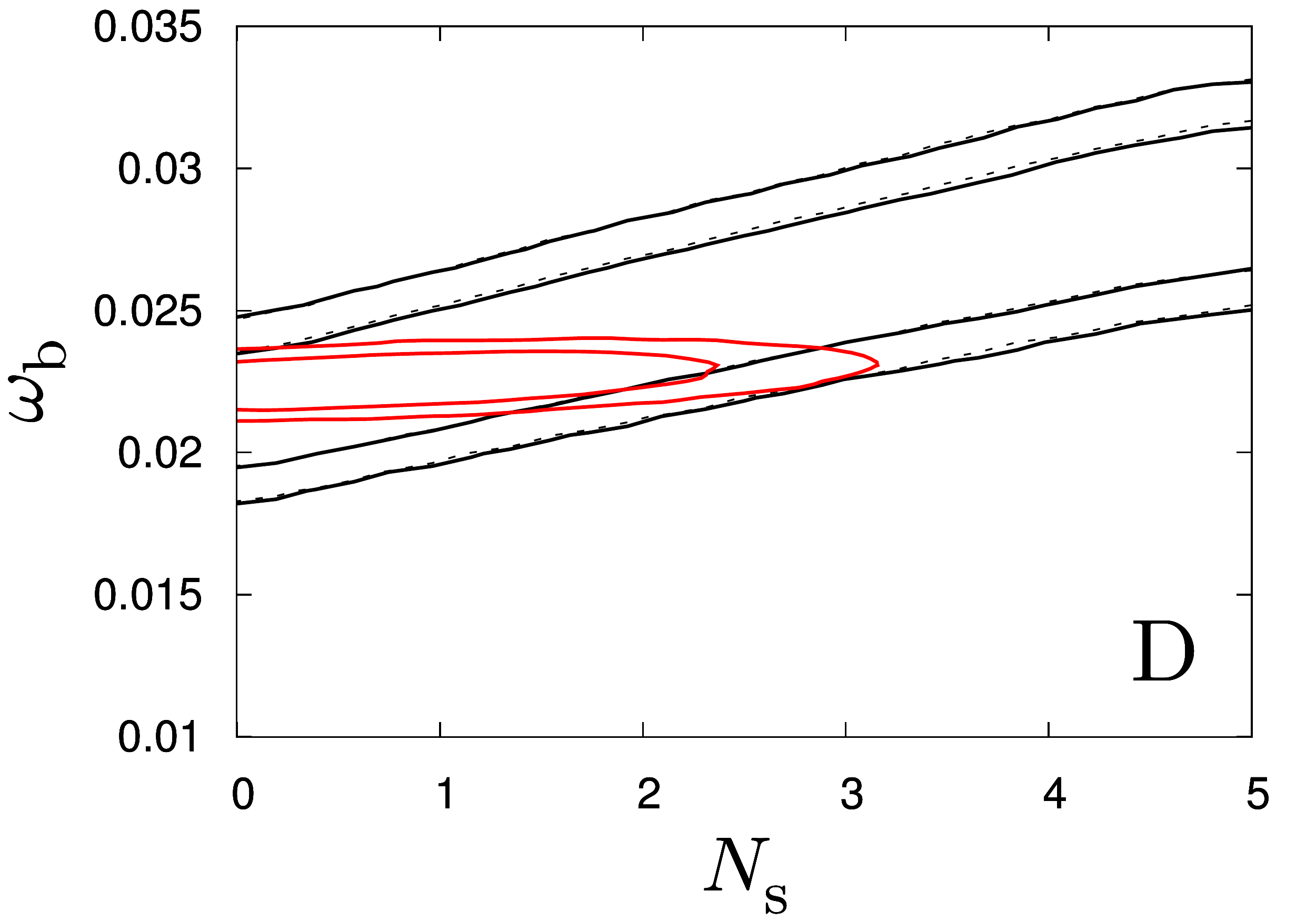}
\includegraphics[height=.48\textwidth,angle=270]{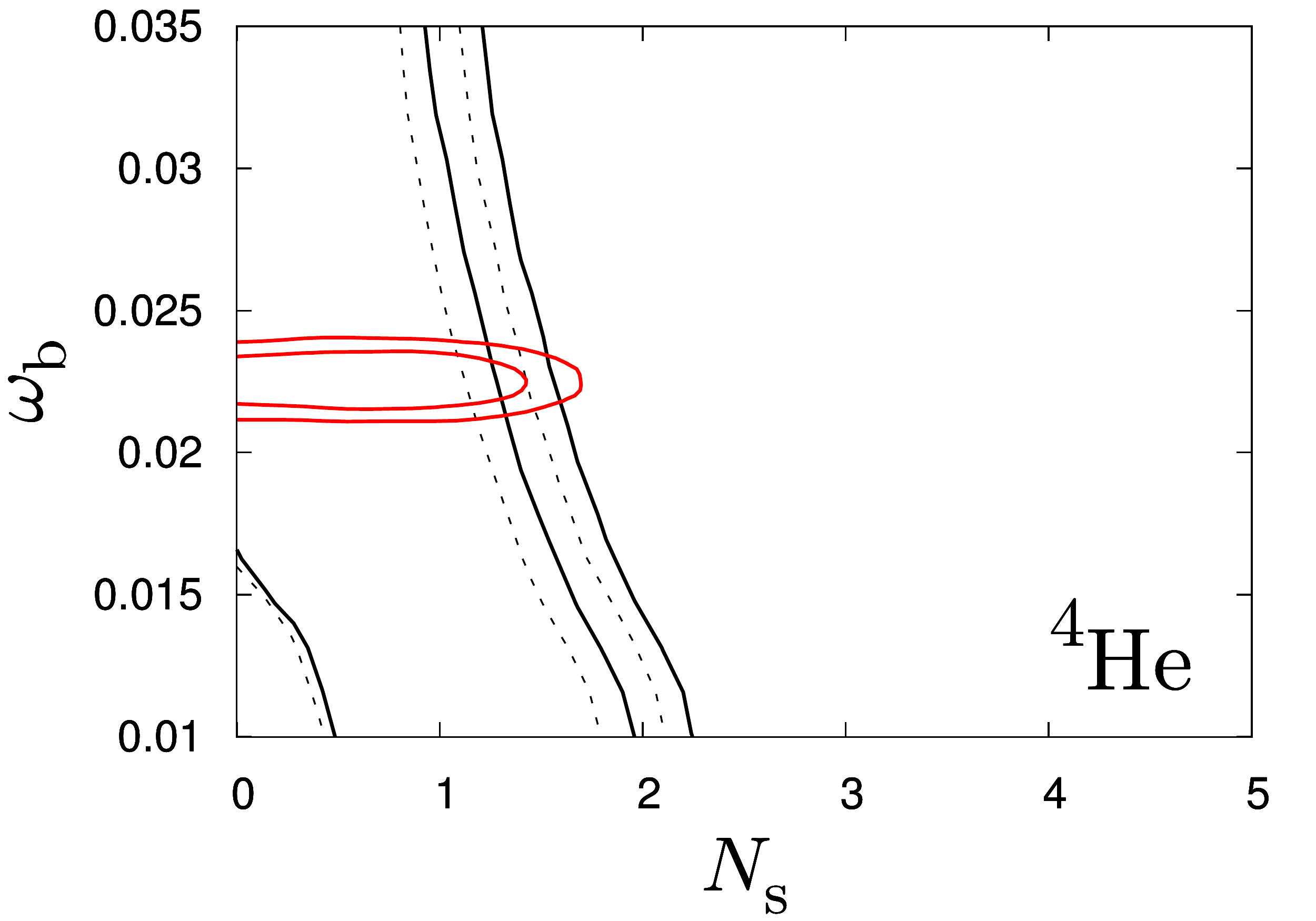}
\begin{center}
\includegraphics[height=.48\textwidth,angle=270]{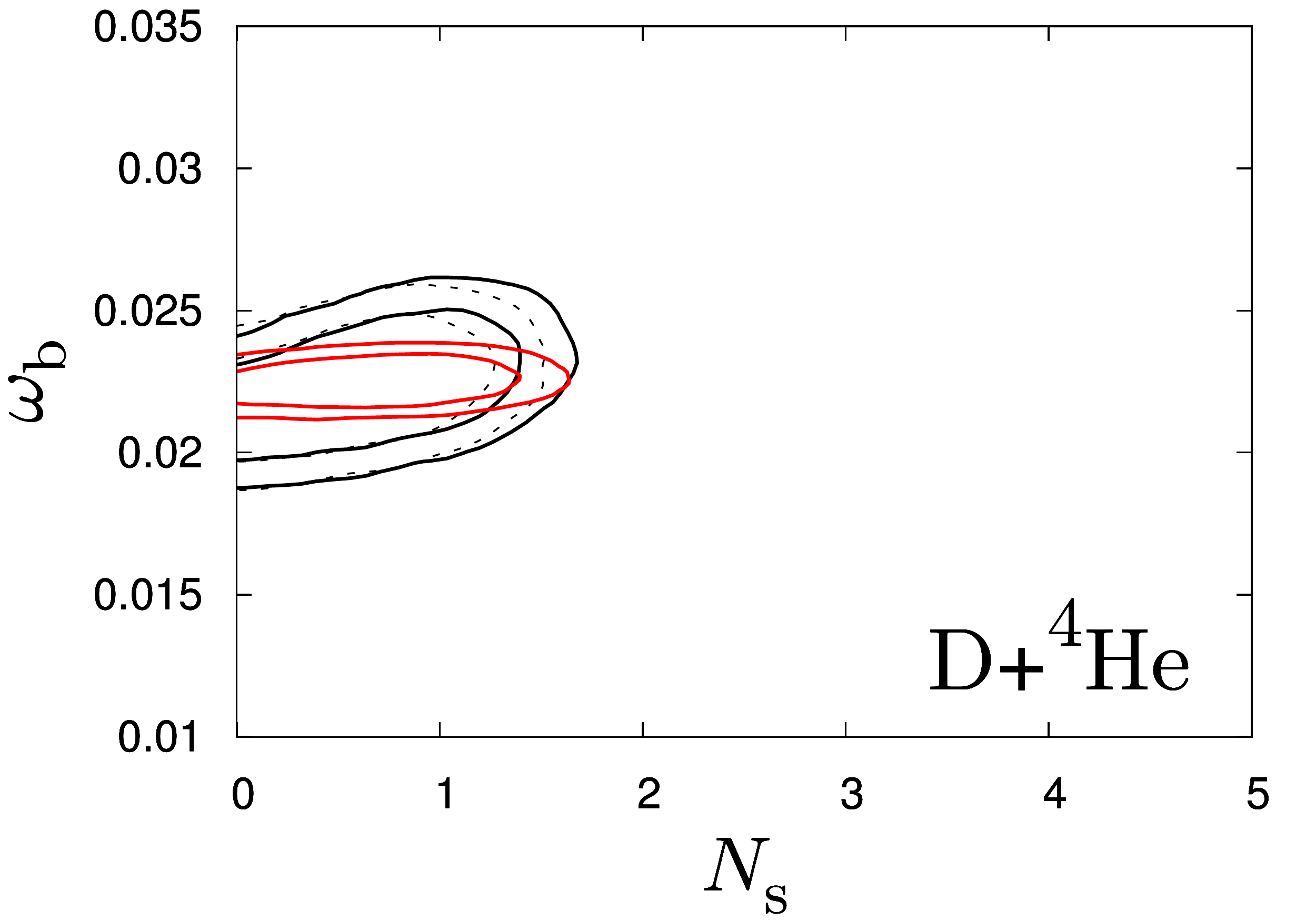}
\end{center}
\caption{2D marginal 90\%- and 99\%-credible regions in
  the $(\omega_{\rm b},N_{\rm s})$-plane, assuming the BBN+sterile
  scenario.  Black lines denote results from elemental abundance
  measurements alone (solid lines assume $\tau_n = 878.5~{\rm s}$,
  dotted lines $885.7~{\rm s}$). The red lines include a
  CMB+LSS prior on $\omega_{\rm b}$.  {\it Top left:} D data.
  {\it Top right:} $^4$He data. {\it Bottom:} D+$^4$He
  data.\label{fig:bbn1}}
\end{figure}

We show the constraints on the $(\omega_{\rm b},N_{\rm s})$
parameter space from our analysis in figure~\ref{fig:bbn1}.  The top
two panels illustrate the well-known fact that the deuterium
abundance constrains mainly the baryon density, whereas constraints
on $N_{\rm s}$ are mostly driven by the helium data. 
However, once we impose the CMB+LSS prior on $\omega_{\rm b}$, even the deuterium data alone
significantly constrain $N_{\rm s}$. This constraint is of interest
in view of the helium data already being systematics-limited.

\begin{figure}[th]
\begin{center}
\includegraphics[height=.48\textwidth,angle=270]{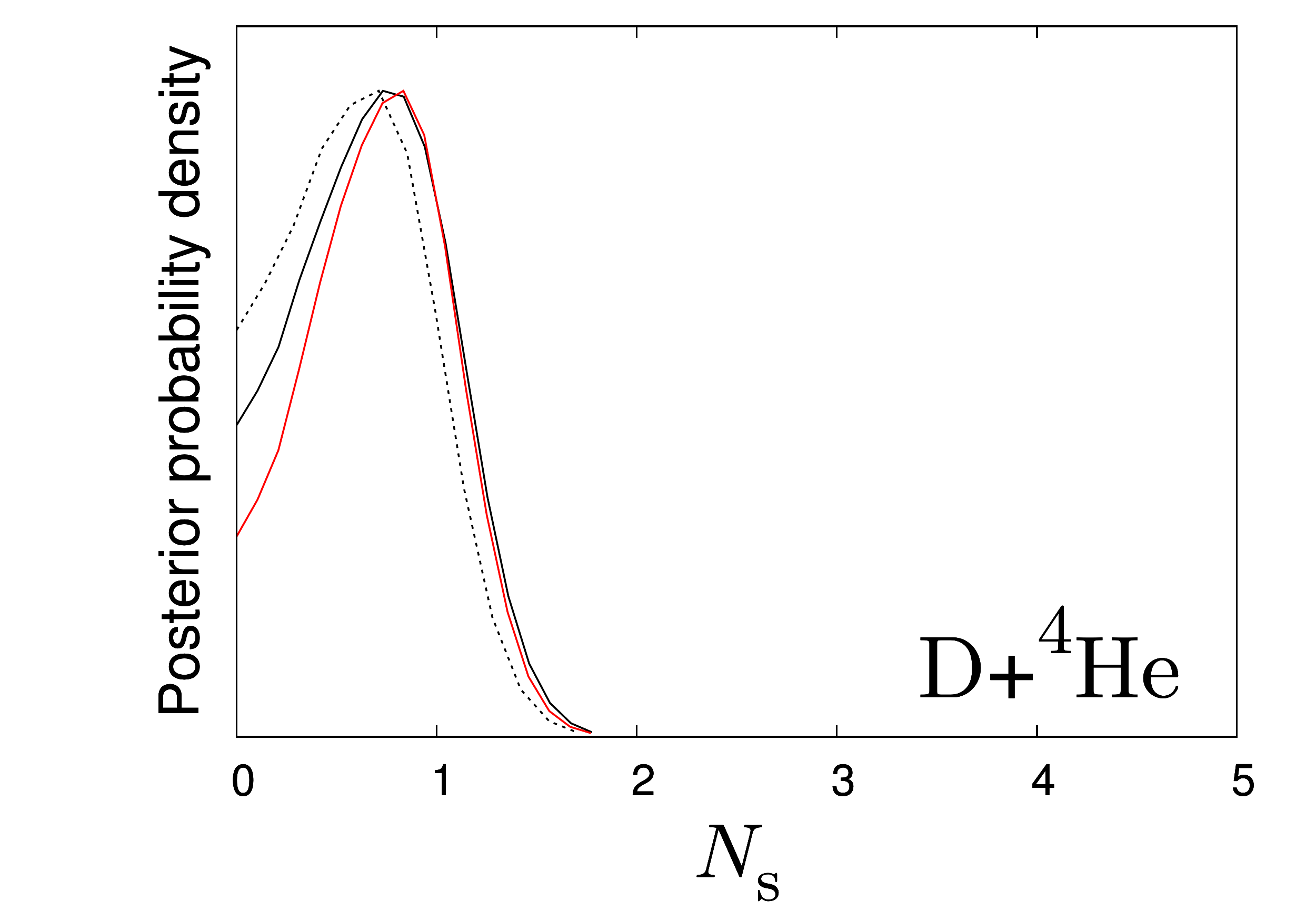}
\end{center}
\caption{1D posterior pdf for $N_{\rm s}$, marginalised over
  $\omega_{\rm b}$, for the BBN+sterile scenario and  D+$^4$He
  data. Same colour/line coding as in figure~\ref{fig:bbn1}.\label{fig:bbn2}}
\end{figure}

For a combined fit of the deuterium and the helium data, the best-fit
value of $N_{\rm s}$ is 0.86, with a 95\%-credible upper limit of
$N_{\rm s} < 1.26$ (or 1.24 if the CMB+LSS prior on $\omega_{\rm b}$
is included).  Using the larger neutron lifetime $\tau_{\rm n}^{\rm
  PDG}$, we obtain slightly lower values: a best-fit of 0.73 and a
95\% upper limit of $N_{\rm s} < 1.14$.  The one-dimensional
marginalised posterior probability densities for $N_{\rm s}$ are shown
in figure~\ref{fig:bbn2}.  As already emphasised in
reference~\cite{Mangano:2011ar}, there is no strong indication for
$N_{\rm s} > 0$ from BBN alone -- $N_{\rm s} = 0$ lies well within the
90\%-credible interval in all cases -- owing to the relatively weak
lower limit on $Y_{\rm p}$.  While one fully thermalised sterile
neutrino species is slightly favoured over $N_{\rm s} = 0$, two fully
thermalised sterile neutrino species are clearly incompatible with the
data in this scenario.

\subsection{Degenerate BBN with sterile neutrinos}

If we imagine for a moment that there were two sterile neutrino
species today, how could this be reconciled with the results of the
previous section?  One possibility would be incomplete thermalisation,
such that the effective $N_{\rm s}$ is smaller than 2; this scenario
could be confirmed if future CMB+LSS data should find $N_{\rm eff} <
5$. Alternatively, the sterile neutrinos could be the decay products
of a heavy particle species between BBN and decoupling.  In this case,
$N_{\rm s}$ could be smaller than 2 at BBN, but equal to or larger
than 2 at decoupling.

A third possibility is the degenerate BBN
scenario~\cite{Kang:1991xa}, in which all standard neutrinos share a
common non-zero chemical potential $\xi$
\cite{Dolgov:2002ab,Pastor:2008ti,Mangano:2010ei}.  The presence of
a chemical potential affects BBN in two ways.  Firstly, it
contributes an additional term to the effective radiation density,
\begin{equation}
  \Delta N_{\rm eff} = \frac{45}{7} \left[ 2  \left(\xi/\pi\right)^2 + \left(\xi/\pi\right)^4 \right].
\end{equation}
Secondly, the degeneracy in the electron neutrinos modifies the
equilibrium neutron-to-proton ratio,
\begin{equation}
  n/p = \exp \left( - \frac{\Delta m}{T} - \xi \right).
\end{equation}
For $\xi \simeq \mathcal{O}(0.1)$, it is the latter effect that is
most relevant for the resulting element abundances: a positive $\xi$
will reduce the number of available neutrons with respect to the $\xi
= 0$ case and thus suppress $Y_{\rm p}$.  This suppression can be used
to circumvent the upper bound on $N_{\rm s}$ found in the previous
section.

\begin{figure}[th]
\begin{center}
\includegraphics[height=.48\textwidth,angle=270]{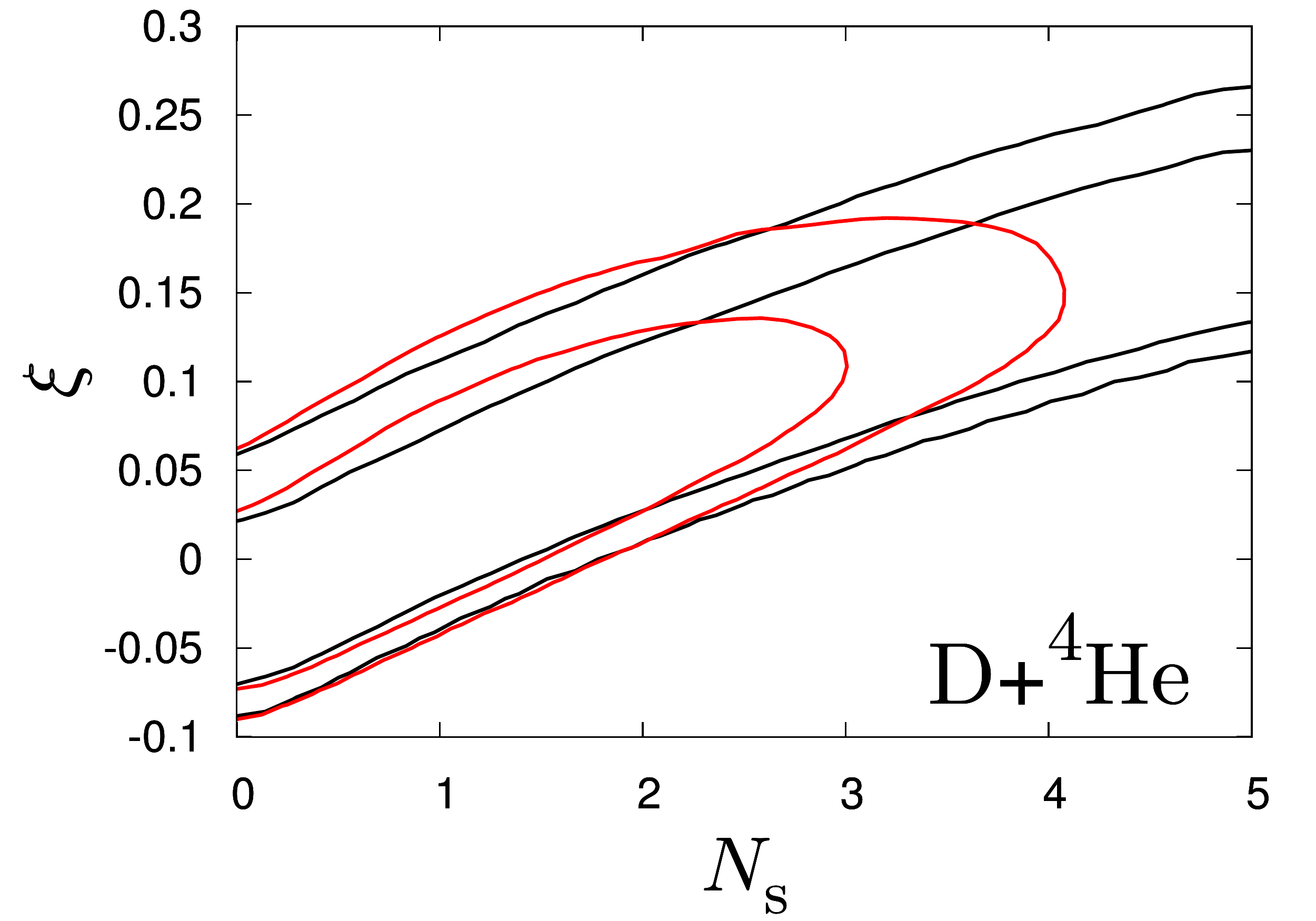}
\end{center}
\caption{2D marginal 90\%- and 99\%-credible regions in the
  $(\xi,N_{\rm s})$-plane marginalised over $\omega_{\rm b}$, assuming
  the degenerate BBN+sterile scenario and $\tau_n = 878.5~{\rm s}$.
  The black lines denote results from D+$^4$He data, while the red
  lines also include a CMB+LSS prior on $\omega_{\rm
    b}$.\label{fig:bbn3}}
\end{figure}

Figure~\ref{fig:bbn3} shows the preferred region in the $(\xi,N_{\rm
s})$-plane in the degenerate BBN scenario. In the 3-dimensional
parameter space $(\omega_{\rm b}, N_{\rm s},\xi)$, the deuterium and
the helium data can only constrain two directions, leaving an
unconstrained one which admits arbitrarily high values of $N_{\rm
s}$.  Only by adding the CMB+LSS prior on $\omega_{\rm b}$ are we
able to obtain any constraint at all: we find a best-fit of $N_{\rm
s} = 0.79$ and a 95\%-credible upper limit of $N_{\rm s} < 2.56$.
Conversely, if we assume $N_{\rm
  s} = 2$, a positive non-zero chemical potential is required: $0.03 <
\xi < 0.14$ (95\%-credible interval), with a best-fit value of $\xi
= 0.064$.

As can be seen from figure~\ref{fig:bbn0}, larger values of $N_{\rm
  s}$ and $\xi$ both lead to a lower prediction for the primordial
$^7$Li abundance.  Taking for instance $N_{\rm s} = 2$, $\xi =
0.066$ and $\omega_{\rm b} = 0.0225$, one obtains $\left[ ^7{\rm
Li}/{\rm
    H}\right]_{\rm p} = 0.372 \times 10^{-10}$ (compared to$\left[
  ^7{\rm Li}/{\rm H}\right]_{\rm p} = 0.460 \times 10^{-10}$ for
$N_{\rm s} = \xi = 0$).  While this is still $3.4$ standard
deviations larger than the measured value, the discrepancy is not
quite as serious as the 4.8 standard deviations one finds in
standard BBN.

\subsection{Combining BBN with CMB+LSS data}
If one assumes that neither $\omega_{\rm b}$ nor $N_{\rm eff}$ changed
between the BBN era and the time of photon decoupling, one could
perform a combined analysis of the CMB+LSS and BBN data in analogy to
the previous literature~\cite{Hannestad:2003xv,Barger:2003zg}.

With the stringent upper bound on the radiation density from the
standard BBN setting, it is clear that the scenarios with one eV-mass
sterile neutrino plus extra massless degrees of freedom considered in
section~\ref{sec:cmblss} will fit the combined data rather badly: for
a 1~eV (2~eV) sterile neutrino, the $\Lambda$CDM+$\Delta N$ model yields
$\Delta \chi^2_{\rm eff} \simeq 9\ (19)$, relative to $\Lambda$CDM. This,
in turn suggests the necessity for further modifications of the
cosmological model: either by dropping the assumption of $N_{\rm eff}$
(or $\omega_{\rm b}$) being constant, or, as seen above, by admitting a nonzero
neutrino chemical potential.

Aside of the massive sterile neutrino scenario, the combination of
CMB+LSS and BBN data can of course also be used to constrain the
commonly considered case of $N_{\rm eff}$ massless degrees of
freedom. We find $N_{\rm eff} = 3.90^{+0.39}_{-0.56}$ at 95\%
credibility, with the standard model expectation of $N_{\rm eff} =
3.046$ outside the 99.5\%-credible region, similar to the limits
reported in the recent work of Hou {\it et al.}~\cite{Hou:2011ec}.

\section{Conclusions\label{sec:discussion}}

In this work we have investigated the effects of eV-mass sterile neutrinos, as
suggested by global interpretations of neutrino oscillation data,
on cosmology.   Such sterile neutrinos can thermalise prior to neutrino decoupling, 
thus contributing to the relativistic energy density in the early universe.
However, while the combination of CMB+LSS and BBN data does appear to prefer 
extra relativistic degrees of freedom at 99.5\% credibility within the $\Lambda$CDM framework,
fully thermalised massive sterile neutrinos in the 1--2~eV mass range necessarily violates
the hot dark matter limit on the maximum neutrino mass.
In terms of the goodness-of-fit,  adding one massless sterile neutrino species
 improves the CMB+LSS fit by \mbox{$\Delta \chi^2_{\rm eff} =-3.16$} relative to standard $\Lambda$CDM, whereas 
 endowing this sterile state with a mass of 1~eV worsens the fit by \mbox{$\Delta \chi^2_{\rm eff} =4.20$}
relative to the same benchmark. Such a scenario is then
excluded or strongly disfavoured.

Nonetheless, while it appears difficult to accommodate eV-mass sterile neutrinos 
within the $\Lambda$CDM framework,  extending the framework with 
modifications in the neutrino sector improves to some extent the consistency 
of sterile neutrinos with precision cosmological data. 
The simplest such modification is to admit even more additional (effectively massless) 
relativistic degrees of freedom, not necessarily fully thermalised. 
Such a scenario improves somewhat the bad effect of the 1~eV sterile neutrino mass 
at the expense of introducing  an additional 1.5 massless species, 
but  the $\chi^2_{\rm eff}$ of the fit is still worse than standard $\Lambda$CDM by
$2.26$ units.  Allowing in addition for a dark energy equation of state $w \neq -1$
further improves the fit: for a 1~eV sterile neutrino, a model with $w=-1.11$ and 
$1.23$ additional massless species in fact fits the data marginally better than standard $\Lambda$CDM ($\Delta \chi^2_{\rm eff} = -0.78$).

Importantly, any model containing eV-mass sterile neutrinos will induce an upward shift in the cold dark matter density
inferred from precision cosmological data.  This shift can have important consequence for, e.g., the SUSY dark matter parameter space.

However, as is well known, increasing the radiation content in the early universe 
can be problematic for BBN.  We find that while standard BBN prefers roughly an extra relativistic
degree of freedom, which we interpret here as a thermalised 1--2~eV sterile neutrino species, 
additional fully thermalised massless species are strongly disfavoured.
Nevertheless,  it is possible to circumvent these BBN constraints 
 with the introduction of a $\nu_e$ chemical potential, which itself could have been
 created by active-sterile oscillations in the early universe.  Thus, the state of affairs can be summarised 
 as follows.  We need additional radiation to reduce the bad effect of hot dark matter on
precision cosmology, and a small neutrino chemical potential to undo the bad
effect of too much radiation on BBN.  In principle, both of these
ingredients can originate from the neutrino sector alone.

In summary, it is not trivial to accommodate a strongly mixed
eV-mass sterile neutrino in cosmology. Additional ingredients are required, such as additional radiation, a neutrino chemical potential, or
a nontrivial $w$ parameter.
In all cases, significant changes in the inferred values of other {\it a priori} unrelated 
cosmological parameters are also incurred, e.g., 
an increase in the cold dark matter density. 
Thus, should the experimental indications
for eV-mass sterile neutrinos become stronger, one must consider a
fairly complex modification of the standard $\Lambda$CDM cosmology.
On the observational side, the upcoming precision measurement of
$\Delta N_{\rm ml}$ by Planck \cite{Perotto:2006rj,Hamann:2007sb}
remains one of the most promising windows to physics beyond the
standard model.

\section*{Acknowledgements}

We acknowledge computing resources from the Danish Center for
Scientific Computing (DCSC). JH acknowledges support from
a Feodor Lynen-fellowship of the Alexander von Humboldt Foundation.
GR acknowledges partial support by the
Deutsche Forschungsgemeinschaft under grants No.\ TR~27 and EXC~153.


\end{document}